%% file: main.tex
\documentclass[10pt,conference]{IEEEtran}
\IEEEoverridecommandlockouts
\usepackage{amsmath,amssymb,amsfonts}
\usepackage{hyperref}
\usepackage{algorithmic}
\usepackage{graphicx}
\usepackage{accessibility}
\usepackage{textcomp}
\usepackage[normalem]{ulem}
\usepackage{array}
\usepackage{cite}

\def\BibTeX{{\rm B\kern-.05em{\sc i\kern-.025em b}\kern-.08em
    T\kern-.1667em\lower.7ex\hbox{E}\kern-.125emX}}
    
\usepackage[svgnames]{xcolor}

\usepackage{tabularx}
\usepackage{booktabs}
\usepackage{multicol}
\usepackage{multirow}
\usepackage{soul}
\usepackage{stfloats}

\newcommand{\hlc}[2][yellow]{{%
    \colorlet{foo}{#1}%
    \sethlcolor{foo}\hl{#2}}%
}

\usepackage{xspace}
\newcommand{\skillmd}{\texttt{SKILL.md}\xspace}

\usepackage{etoolbox}
\makeatletter
\patchcmd{\@makecaption}
  {\scshape}
  {}
  {}
  {}
\makeatother

\graphicspath{{Figures/}}

\usepackage{wrapfig}

\newcommand{\papertitle}{From Anatomy to Smells: An Empirical Study of SKILL.md in Agent Skills}

\hypersetup{
    colorlinks=true,
    linkcolor=Blue,
    filecolor=magenta,      
    urlcolor=Blue,
    citecolor=Blue,
    pdftitle={\papertitle}
}

\usepackage[many]{tcolorbox}    	%
\definecolor{main}{HTML}{31363F}    %
\definecolor{sub}{HTML}{EEEEEE}     %
\tcbset{
    sharp corners,
    colback = white,
}  
\newtcolorbox{boxH}{
    colback = sub, 
    colframe = main, 
    boxrule = 0pt, 
    leftrule = 5pt %
}
\newtcolorbox{boxA}{
    fontupper = \bf,
    boxrule = 1pt,
    colframe = main %
}
\usepackage{subfiles} %

\usepackage{colortbl}

\newcommand{\categoryrow}[0]{
\rowcolor{LightBlue}
}

\begin{document}
\bstctlcite{IEEEexample:BSTcontrol}

\title{\papertitle}

\author{
\IEEEauthorblockN{David Boram Hong}
\IEEEauthorblockA{
\textit{University of California, Irvine}\\
Irvine, USA \\
dbhong@uci.edu}
\and
\IEEEauthorblockN{Aaron Imani}
\IEEEauthorblockA{
\textit{University of California, Irvine}\\
Irvine, USA \\
aaron.imani@uci.edu}
\and
\IEEEauthorblockN{Iftekhar Ahmed}
\IEEEauthorblockA{
\textit{University of California, Irvine}\\
Irvine, USA \\
iftekha@uci.edu}
}

\maketitle

\thispagestyle{plain}
\pagestyle{plain}

\begin{abstract}
Agent Skills provide on-demand domain knowledge to LLM agents without requiring model retraining. Each Agent Skill is defined by a mandatory \texttt{SKILL.md} file containing metadata and an unstructured Markdown body whose contents are left entirely to the skill author. Despite the rapid adoption of Agent Skills, little is known about how these files are authored or whether existing authoring guidelines are followed in practice. In this paper, we present the first systematic study of \texttt{SKILL.md} files as a software artifact. We qualitatively analyze 238 real-world skills and derive a taxonomy of 13 higher-level and 44 lower-level semantic components. We then conduct a multivocal literature review of 29 sources to identify best practices for authoring \texttt{SKILL.md} files and introduce \emph{skill smells} as violations of these practices. Finally, we develop an automated detector and apply it to real-world skills, finding that over 99\% of \texttt{SKILL.md} files contain at least one skill smell, and once introduced, skill smells rarely disappear as skills evolve. These findings reveal a substantial gap between recommended and actual authoring practices, motivating the development of automated techniques to remediate skill smells while increasing developer awareness of this emerging quality issue.
\end{abstract}

\begin{IEEEkeywords}
agent skill, SKILL.md
\end{IEEEkeywords}

\section{Introduction}

\subfile{Sections/intro}

\section{Related Work} \label{related-work}

\subfile{Sections/related-work}

\section{Methodology} \label{method}
\subfile{Sections/methodology}

\section{Results} \label{results}
\subfile{Sections/results}

\section{Discussion} \label{discussion}

\subfile{Sections/implications}

\section{Threats to Validity} \label{threats}
\subfile{Sections/threats}

\section{Conclusion} \label{conclusion}
\subfile{Sections/conclusion}

\bibliographystyle{IEEEtran}
\bibliography{IEEEabrv, references}

\end{document}

%% file: Sections/intro.tex
With the recent growth of autonomous software agents, such as Claude Code~\cite{claude-code} and GitHub Copilot~\cite{copilot}, that use Large Language Models (LLMs) to perform Software Engineering (SE) tasks, the ability to provide agents with task-specific and project-specific knowledge has become increasingly important. Agent Skills\footnote{We use ``skill'' and ``Agent skill'' interchangeably.}~\cite{agent-skills} address this need by packaging procedural knowledge together with organization-, team-, and user-specific context into portable, version-controlled artifacts that agents can retrieve on demand. As evidenced by the emergence of public skill marketplaces containing more than 100,000 skills~\cite{skills-sh,agent-skills-sh}, skills have become a widely adopted mechanism for adapting general-purpose agents to specialized tasks. 

Originally introduced by Anthropic~\cite{agent-skills}, the Agent Skill Specification defines a skill as a directory containing a required \texttt{SKILL.md} file and optional scripts, references, and assets~\cite{skill-specification}. The \texttt{SKILL.md} file consists of frontmatter, a metadata header in YAML format \cite{yaml} with a set of predefined supported fields \cite{skill-specification}, which helps agents identify and load relevant skills, and a Markdown body containing skill instructions. Figure~\ref{fig:rq3-smell-examples} shows an excerpt from a \texttt{SKILL.md} body alongside two excerpts from frontmatter sections. Beyond these structural requirements, skill authors are free to include any content that helps agents perform the target task effectively. 

The largely unconstrained nature of \texttt{SKILL.md} body and content create the potential for substantial variation in skill quality. Because skills directly contribute to the context provided to agents, poorly written skills may introduce irrelevant information, omit important guidance, or present instructions in ineffective ways, ultimately resulting in suboptimal agent behavior. Such effects may impact not only task performance but also resource consumption by increasing context length and reducing the usefulness of retrieved information. Despite the rapid adoption of agent skills, prior research has primarily focused on their effectiveness~\cite{li2026skillsbenchbenchmarkingagentskills} and security implications~\cite{beurerkellner2026technicalreportexploringemerging,guo2026skillprobesecurityauditingemerging,liu2026agentskillswildempirical,schmotz2025agentskillsenablenew}, leaving their authoring practices largely unexplored. Understanding this gap is particularly important because skills may increasingly be used as training data for future models and may themselves be generated by LLMs, creating the potential for poor authoring practices to propagate through both training and generation pipelines.

The history of SE shows that widely used textual artifacts tend to accumulate quality problems when they evolve without clear quality guidelines. For example, prior work found that commit messages, which are also free-form text describing a code change, were often incomplete, uninformative, or even empty~\cite{Dyer2013}, and that poor-quality commit messages were associated with software defects~\cite{li2023CM-Matters}. Because Agent Skills are likewise largely unconstrained in both structure and content, we posit that they represent an emerging software quality concern for agentic systems. Since Agent Skills are still in the early stages of adoption, there is a unique opportunity to understand current authoring practices and identify problematic patterns before they become widespread and entrenched. Hence, we conduct the first large-scale study of \texttt{SKILL.md} quality, best practices, and quality deficiencies in real-world agent skills to understand what information they contain, characterize the semantic components of their bodies.

\textbf{RQ1: What are the semantic components of SKILL.md files in practice?}

The next step toward assessing \texttt{SKILL.md} quality is understanding how their semantic components should be authored. Although practitioners and organizations have proposed best practices for writing \texttt{SKILL.md} files~\cite{best-practices-google}, this guidance is scattered across disparate sources, making it unclear which semantic components are covered and which lack guidance. We therefore systematically collect and map existing best practices to the semantic components they govern. This establishes a notion of what constitutes well-authored \texttt{SKILL.md} files and provides the foundation for identifying quality deficiencies through violations of these practices. %

Prior work in SE has shown that violations of established best practices often manifest as software quality problems. For example, violations of coding best practices, commonly referred to as \textit{code smells}, have been associated with increased fault-proneness, reduced maintainability, and lower software quality~\cite{palomba2018,Saboury2017}. Motivated by this line of work, we define \textit{skill smells} as violations of best practices for authoring \texttt{SKILL.md} files. Therefore, in our second research question, we investigate the best practices for authoring \skillmd files, the semantic components they govern, and the resulting skill smells that arise when these practices are violated.

\textbf{RQ2: What best practices exist for authoring \skillmd, and what skill smells emerge from their violation?}

Because agents rely on skills to obtain task-specific context and procedural guidance, the quality of a \texttt{SKILL.md} file directly affects the quality of the information supplied to the agent. Similar to code smells, \textit{skill smells} may introduce unnecessary complexity, reduce the effectiveness of the guidance provided, and degrade the quality of the context on which agents rely for reasoning and decision-making. For example, unnecessary content in a skill may consume valuable context-window capacity and distract the agent from the task at hand~\cite{gao2026skillreduceroptimizingllmagent}. However, before developing techniques to detect or mitigate skill smells, it is first necessary to understand how prevalent they are in real-world Agent Skills. Measuring their prevalence provides the first empirical evidence of the scale of the problem and helps prioritize future quality-improvement efforts. This leads to our next research question.

\textbf{RQ3: How prevalent are skill smells in real-world \texttt{SKILL.md} files?}

Identifying skill smells is only the first step toward understanding their impact on the agent-skill ecosystem. An important follow-up question is whether they are corrected over time or instead persist as skills evolve. Prior SE research has shown that quality characteristics often deteriorate throughout software evolution. For example, commit-message quality decline as software evolves~\cite{li2023CM-Matters}, the prevalence of code smells tends to increase in open-source repositories~\cite{Ahmed2015}. Such longitudinal analyses help determine whether quality issues are naturally mitigated through maintenance activities or become increasingly entrenched over time. Understanding this evolution is particularly important for skill smells because persistent quality issues may continue to affect the context and guidance provided to agents. Moreover, if skill smells are rarely corrected once introduced, they strengthen the case for automated quality-assurance tools and improved authoring practices. We therefore investigate how the prevalence of skill smells evolve throughout the lifetime of real-world skills.

\textbf{RQ4: How does the prevalence of skill smells evolve over a skill's lifetime?}

\noindent Overall, our study contributes in the following ways:

\begin{itemize}
    \item We are the first to characterize the anatomy of \texttt{SKILL.md} files beyond frontmatter and body, providing a taxonomy of semantic components present in real-world agent skills.
    \item We systematically identify best practices proposed for \texttt{SKILL.md} effectiveness and map each practice to the component(s) it governs.
    \item We introduce the concept of \textit{skill smells} in \texttt{SKILL.md} files that could impair agent performance in the agentic paradigm and propose \textbf{S}kill \textbf{S}mell \textbf{D}etector (\textbf{SSD}) to automatically identify them.
    \item We empirically analyze how the prevalence of skill smells evolves over time and demonstrate that most skill smells have become increasingly prevalent across our dataset.
\end{itemize}

The remainder of this paper is structured as follows. In \hyperref[related-work]{Section II}, we review related research relevant to our study. \hyperref[method]{Section III} describes our methodology for addressing all research questions. \hyperref[results]{Section IV} presents the results of our experiments. \hyperref[discussion]{Section V} discusses the potential effects of skill smells on agents according to the literature. In \hyperref[threats]{Section VI}, we outline potential threats to the validity and describe the steps taken to mitigate them. Finally, \hyperref[conclusion]{Section VII} concludes the paper and suggests directions for future work.

%% file: Sections/related-work.tex
\subsubsection*{\textbf{Automated Construction and Evolution of Agent Skills}}

Recent work has explored methods for automatically constructing and evolving agent skills. CoEvoSkills~\cite{zhang2026coevoskillsselfevolvingagentskills} enables agents to autonomously create and refine multi-file skill packages, while SkillClaw~\cite{ma2026skillclawletskillsevolve} continuously updates skills by mining agent trajectories and converting recurring behaviors into skill revisions. These studies focus on how skills can be generated and evolved automatically. In contrast, we focus on the quality of \texttt{SKILL.md} files, the best practices that govern them, and how quality deficiencies persist over time.

\subsubsection*{\textbf{Empirical Studies of Agent Skills}}

Several studies have examined agent skills, characteristics, and usage. Ling et al.~\cite{ling2026agentskillsdatadrivenanalysis} conducted a large-scale analysis of 40,285 skills from skills.sh~\cite{skills-sh}, reporting that \texttt{SKILL.md} files are typically compact enough to fit alongside planning context and tool schemas within an agent's context window. They further found that SE was the most common application domain, accounting for 54.7\% of the skills analyzed. Gao et al. \cite{gao2026skillreduceroptimizingllmagent} analyzed %
paragraph-level items from 90 \texttt{SKILL.md} files and classified them into five pre-defined categories: core rules, background, examples, templates, and redundant content to optimize skill compression. In contrast, our goal is not to categorize paragraphs for optimization, but to understand \texttt{SKILL.md} files as a software artifact. Hence, rather than relying on predefined categories, we inductively derive a taxonomy of semantic components from real-world skills as the foundation for identifying best practices and defining skill smells.

\subsubsection*{\textbf{Effectiveness and Optimization of Agent Skills}}

Researchers have investigated the efficiency of agent skills. SkillsBench~\cite{li2026skillsbenchbenchmarkingagentskills} evaluated agent skills across 86 tasks spanning 11 domains and found that curated skills can provide substantial, but highly variable, performance improvements. SWE-Skills-Bench~\cite{han2026sweskillsbenchagentskillsactually} examined the effectiveness of skills in SE tasks and found that many skills provide limited benefit, while some can even degrade performance when their guidance conflicts with project-specific context. Other work has focused on optimizing skills after they are created. SkillReducer~\cite{gao2026skillreduceroptimizingllmagent} improves token efficiency by reducing unnecessary skill content, while SkillMOO~\cite{gong2026skillmoomultiobjectiveoptimizationagent} formulates skill optimization as a multi-objective search problem balancing effectiveness and cost.

Collectively, prior work has studied how agent skills are used, generated, evolved, and optimized. However, little attention has been paid to understanding the semantic structure of \texttt{SKILL.md} files, identifying best practices for their authoring, characterizing quality deficiencies, and studying how such deficiencies persist over time. Our work addresses these gaps through a large-scale empirical study of \texttt{SKILL.md} content, best practices, and skill smells.

%% file: Sections/methodology.tex
\begin{figure*}[t!]
  \centering
  \includegraphics[width=0.85\textwidth]{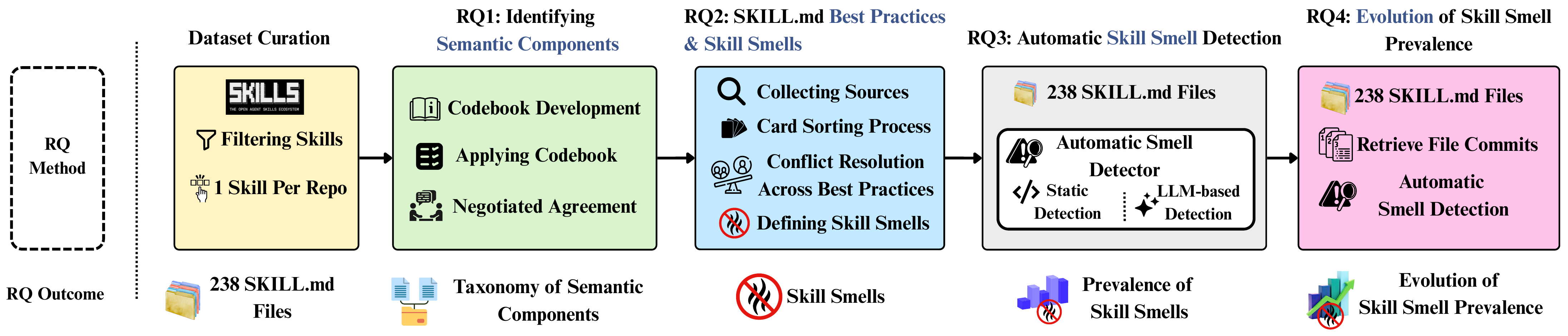}
  \caption{Overall pipeline of the study.}
  \label{fig:overall}
\end{figure*}

We begin by describing the skills sample dataset used throughout our analyses, and then describe our methodology, as illustrated in Figure \ref{fig:overall}. 

\textbf{Dataset Curation.} Following previous work in agent skills studies \cite{holzbauer2026context, chen2026skvmrevisitinglanguagevm, ling2026agentskillsdatadrivenanalysis}, we used publicly available skills from the skills.sh marketplace \cite{skills-sh}. Specifically, we used a public dump of agent skills from this marketplace \cite{open-skills}, which contains 133,149 skills from 8,808 publishers across 13,460 repositories. From this dataset, we applied a multi-step filtering and sampling procedure following mining best practices \cite{dabic2021}, based on language \cite{liu2026agentskillswildempirical}, adoption signals \cite{dabic2021, liu2026agentskillswildempirical}, and repository diversity. 
Specifically, we first removed all non-English skills using fast--langdetect full model \cite{joulin2016bag, joulin2016fasttext} consistent with prior work \cite{liu2026agentskillswildempirical}, retaining 123,319 English skills and discarding 9,830 non-English ones. Second, to focus on skills with meaningful real-world adoption, we retained skills with at least 10 weekly downloads (58,311 skills), then further restricted the sample to the top 10th percentile by repository star count \cite{dabic2021}, corresponding to a minimum of 8,000 stars (5,410 skills). Third, we deduplicated skills by name, retaining the version with the highest star count, resulting in 4,061 unique skills. 
Next, since multiple skills can belong to the same repository, to avoid over-representing any single repository \cite{napaggan2013}, we grouped skills by repository and owner and randomly sampled one skill per repository, producing a set of 241 skills \cite{dabic2021}. We sampled one skill per repository because it represented the minimum number of skills present across all repositories; sampling a larger fixed number would have left some repositories underrepresented or excluded entirely, resulting in an unbalanced sample.
Out of the 241 sampled skills, we were unable to retrieve 3 due to issues such as mismatched skill names, non-default branch hosting, or post-publication deletion. Our final sample therefore consists of 238 \texttt{SKILL.md} files.

\subsection{\textbf{RQ1: Semantic Components of \texttt{SKILL.md} Body}} \label{method:rq1} In order to identify the semantic components that construct the body of \texttt{SKILL.md} files, we conducted a qualitative analysis on the \texttt{SKILL.md} files. 
Following prior work that uses headings as the primary unit of analysis in Markdown files \cite{Prana2019}, we analyzed \texttt{SKILL.md} bodies at the H2 heading level. We chose this granularity because H1 conventionally captures the document title and only occurs once per Markdown document \cite{gitlab-handbook}, making H2 the coarsest meaningful structural unit for content analysis.

We conducted a two-stage qualitative content analysis following established practices~\cite{tafreshipour2025}. In the first stage, codebook development, two authors independently analyzed a randomly selected sample of 128 \texttt{SKILL.md} files from a population of 238 files (90\% confidence level, 5\% margin of error) at the H2-heading level to identify fine-grained semantic components. The authors then compared their codes and resolved disagreements through negotiated agreement~\cite{corbin2015basics,strauss1998basics}. Components identified with different terminology were reconciled, and components proposed by a single author were discussed and retained or discarded by consensus. To address semantic overlap among the resulting components, the authors reorganized them into hierarchical categories: lower-level components, and a higher-level category that groups the lower-level components. This produced a hierarchical taxonomy of \texttt{SKILL.md} semantic components.
In the second stage, both authors independently applied the finalized codebook to the remaining 110 \texttt{SKILL.md} files. %
The resulting taxonomy and component distribution are presented in \hyperref[results:rq1]{Section IV}.

\subsection{\textbf{RQ2: \texttt{SKILL.md} Best Practices \& Skill Smells}} \label{method:rq2}

To the best of our knowledge, no systematic literature review or substantial body of scientific work exists on best practices for \texttt{SKILL.md} authoring. Therefore, we conducted a Multivocal Literature Review (MLR) \cite{Rodney1991}.
We followed an approach similar to prior work \cite{li2024omg} by using Google Search \cite{best-practices-google} to retrieve the top 50 online sources containing the terms ``agent skills'' or ``\texttt{SKILL.md}'' in combination with ``best practices.'' Two of the authors independently manually reviewed all retrieved sources for relevance, discarding those that were not text-based, such as video tutorials, and those that contained the query keywords but did not discuss best practices for writing \texttt{SKILL.md} files, such as a blog post by AGENSI \cite{agensio} and then through negotiated agreement~\cite{corbin2015basics,strauss1998basics} finalized 29 sources. 

Next, two authors independently extracted and consolidated candidate best practices from the selected sources using a card-sorting process~\cite{ZIMMERMANN2016137}. Similar recommendations expressed using different wording were grouped together, and disagreements regarding grouping or interpretation were resolved through negotiated agreement~\cite{corbin2015basics,strauss1998basics}. We observed cases where recommendations conflicted across sources. To resolve such conflicts, we applied majority voting among the sources that discussed the recommendation: a best practice was retained if it was supported by a strict majority and discarded in the event of a tie. For instance, if a best practice was discouraged in two sources but encouraged in three, it was retained as a best practice. 

After compiling the final set of best practices, we derived skill smells by inverting each best practice, defining a \textit{skill smell} as an anti-pattern in \texttt{SKILL.md} files that violates the corresponding best practice.
Two authors performed this inversion jointly, discussing the precise wording and scope of each resulting smell to ensure it captured a clear and detectable violation of its corresponding best practice. For example, the best practice recommending that \texttt{SKILL.md} content remain high-level, with context-specific guidance delegated to reference documents, was inverted into the \textit{Undelegated Details} skill smell, defined as containing low-level information and specific guidance that is not delegated to reference documents or scripts. This process resulted in defining 26 skill smells, which we present in Section \ref{results:rq2}.

\subsection{\textbf{RQ3: Automated Detection of Skill Smells}} \label{method:rq3}

In RQ2, we introduced 26 skill smells corresponding to violations of \texttt{SKILL.md} authoring best practices. However, the extent to which each skill smell occurs in practice remains unknown. RQ3 therefore investigates their prevalence across our sample of 238 \texttt{SKILL.md} files. %
Since we are the first to introduce skill smells, we sought a scalable means of identifying them in practice. To this end, we developed an automated skill smell detection tool, Skill Smell Detector (SSD), that serves a dual purpose: measuring the prevalence of skill smells in our sample, and assisting skill authors in detecting smells as they write \texttt{SKILL.md} files, thereby mitigating the potential impact of skill smells on downstream tasks discussed in Section~\ref{discussion}.

We automated the detection process, beginning with the construction of a ground truth dataset. To this end, we randomly sampled 53 \texttt{SKILL.md} files (confidence level 90\%, margin of error 10\%), and two authors independently reviewed each file for occurrences of any of the 26 skill smells. Following the same procedure as RQ1, the authors then held a negotiated agreement session to resolve disagreements, discussing mismatched cases and reaching consensus. This yielded a labeled ground truth dataset of skill smell occurrences across the 53 files. In Section~\ref{results:rq3}, we present Table~\ref{table:skill-smell-freq} that presents the distribution of skill smells identified across the 53 manually labeled files (Validation Dataset).

Skill smells fall into two categories based on the type of reasoning required for their detection. The first category, which we term \textit{Statically Detectable} smells, can be identified using rule-based scripts without any semantic reasoning. For example, the \textit{Oversized \texttt{SKILL.md}} smell, defined as a \texttt{SKILL.md} file exceeding 5,000 words, can be verified with a simple word count script. Therefore, we developed python scripts to detect skill smells within this category. In total, five smells were \textit{Statically Detectable}.
The second category, which we term \textit{Semantically Detectable} smells, cannot be identified through static analysis alone and instead require contextual reasoning over the content of the file. In total, there were 21 \textit{Semantically Detectable} smells. We present the skill smells that fall into each of these two detection types in Table~\ref{table:skill-smell-freq}.

Given the demonstrated strength of LLMs in simulating human judgment on software engineering artifacts \cite{wang2025, ahmed2025llmsreplacemanualannotation}, we employed LLMs to detect \textit{Semantically Detectable} skill smells.
Since open-source LLMs have demonstrated competitive performance compared with proprietary models on software engineering tasks \cite{imani2025omega, zhong2025-larger}, we opted for an open-source LLM to detect semantically detectable skill smells. Inference was performed on a Linux server equipped with an NVIDIA A6000 GPU with 48GB of VRAM. We selected Qwen3.6-27B, which has achieved superior performance across several reasoning benchmarks \cite{qwen3.6-27b}, as our base model. However, deploying it in full precision exceeded our hardware constraints. To address this, we leveraged Activation-aware Weight Quantization (AWQ) \cite{MLSYS2024_42a452cb}, which has been shown to introduce minimal degradation in model performance and has demonstrated strong results on software engineering tasks such as commit message generation \cite{imani2025omega}. Specifically, we used a 4-bit AWQ quantized variant of Qwen3.6-27B available on HuggingFace \cite{Qwen3.6-27B-AWQ-INT4}.

Prior work has shown that structured output formats such as JSON improve LLM-based classification \cite{tam-etal-2024-speak}, and that providing examples enhances LLM annotation performance on software engineering artifacts \cite{ahmed2025llmsreplacemanualannotation}. Drawing on these findings, we prompted the model in thinking mode to enable extended reasoning capabilities \cite{qwen3.6-27b}. We chose the sampling parameters per the recommendation of the model card for ``Thinking mode for general tasks'' \cite{Qwen3.6-Card}. The prompt consisted of an introduction followed by a per-smell section for each of the 21 Semantically Detectable smells. The introduction informed the model about the nature of \texttt{SKILL.md} files, the concept of skill smells, and its role as a skill smell detector. For each skill smell, we provided its definition, a positive example showing a \texttt{SKILL.md} excerpt that exhibits the smell, and a matched negative example showing the same excerpt without the smell, to help the model distinguish between the presence and absence of each smell. We provide the prompt in our supplementary material \cite{SupplementaryMaterials}. 
We evaluated the effectiveness of our LLM-based semantic skill smell detector by measuring precision, recall, and F1-score on the manually curated ground truth dataset, with results reported in Section~\ref{results:rq3}. Finally, by combining the static and semantic detectors, we built an end-to-end pipeline that takes a \texttt{SKILL.md} file as input and outputs the list of skill smells it contains. We then applied SSD to the remaining 238 \texttt{SKILL.md} files in our sample to measure the prevalence of each skill smell.

\subsection{\textbf{RQ4: How Skill Smell Prevalence Evolves Over Time}} \label{method:rq4}

To understand how skill smells persist over a skill's life-cycle, we mined the complete commit history of each of the 238 \texttt{SKILL.md} files in our sample, yielding 1,295 commit records. As the purpose of this RQ is to understand the persistence of skill smells, we excluded all skills with a single commit (did not change after the initial commit), leaving 142 \texttt{SKILL}.md files, consisted of 1,199 commit records. We treated each commit as a distinct version of the file and applied our automated Skill Smell detector (introduced earlier in Section \ref{method:rq3}) to every version to identify the skill smells it contained. For each file, we then ordered its versions chronologically and, for each skill smell category, calculated the version-by-version sequence of the smell's presence or absence.

Then, we projected these version sequences onto a common weekly timeline to characterize how smell prevalence evolves over time. We defined the observation window as the interval spanning the earliest and latest commit dates across all 142 files (October 16, 2025 to June 13, 2026) and partitioned it into 35 consecutive, non-overlapping weeks. Each observation window contained the earliest commit for each file. A file is considered to \emph{exist} from the week of its initial commit onward. For each week, we determined the state of every existing file using a last-observation-carried-forward scheme: a file's smell profile in a given week is taken from its most recent commit on or before the end of that week. This reflects that a \texttt{SKILL.md} file retains the smells of its latest version until a subsequent commit modifies it, so that weeks without a new commit inherit the preceding state. Given these weekly file states, we computed, for each smell category and each week, its prevalence: the proportion of files existing in that week whose current version exhibited the smell.

%% file: Sections/results.tex
\subsection*{\textbf{RQ1: Semantic Components of \skillmd Body}} \label{results:rq1}

\begin{table*}[ht!]
\caption{Taxonomy of \skillmd Body Semantic Components (Frontmatter is not included as a component since it is not part of the \skillmd body and its structure is enforced by the guidelines \cite{skill-specification})}

\label{table:body-semantic-component}
\centering
\resizebox{0.85\textwidth}{!}{%
\begin{tabular}{lp{10cm}}
\toprule
\textbf{Lower-Level Semantic Component} & \multicolumn{1}{c}{\textbf{Definition}} \\
(\# Occurrence/Total) & \\
\midrule
\categoryrow
\multicolumn{2}{c}{\textbf{Task (74.4\%)}} \\
Steps Instruction (49.2\%) & %
Instructions that decompose a workflow into manageable steps for the agent.\\
Subtask Instruction (24.0\%) & %
A self-contained procedure for accomplishing a specific task.\\
Error Handling (18.5\%) & %
Instructions for handling errors encountered during task execution.\\
Environmental Variation (10.1\%) & %
Instructions for handling different execution environments (e.g., Windows vs. Linux).\\

\midrule
\categoryrow
\multicolumn{2}{c}{\textbf{Introduction (63.5\%)}} \\
Prerequisites (31.1\%) & Conditions that must be fulfilled to proceed with the skill. \\
Skill Trigger (23.1\%) & %
Conditions under which the skill should or should not be invoked.\\
Skill Summary (21.0\%) & Overview of what the skill tries to accomplish. \\
Quick Start (8.8\%) & %
Commands for quickly using the skill.\\
Skill Directory Format (8.0\%) & Description on how the folder that contains \texttt{SKILL.md} is structured. \\
Authentication (4.6\%) & %
Authentication and security requirements.\\

\midrule
\categoryrow
\multicolumn{2}{c}{\textbf{References (57.1\%)}} \\
Reference Files (31.1\%) & %
Descriptions of supplementary documents that agents may load on demand.\\
Commands (15.1\%) & %
CLI commands available to the agent.\\
Domain Knowledge (13.9\%) & %
Domain-specific knowledge relevant to the skill.\\
Tips (13.5\%) &  Advisory guidance for the agent. \\
Related Skills (8.0\%) & Information on other skills that may be relevant to the task that the agent is performing. \\
\multirow{2}{*}{Utility Scripts (5.9\%)} & Scripts that are provided for the agent to use rather than leaving one to generate one by itself. \\
Standards (1.3\%) & %
References to relevant industry standards or specifications.\\

\midrule
\categoryrow
\multicolumn{2}{c}{\textbf{Principles (42.0\%)}} \\
Rules (39.1\%) & Rules for the agent to follow. \\
Generation Quality (5.5\%) & %
Quality requirements for generated artifacts.\\
Instruction Hierarchy (1.3\%) & %
Guidance on which instruction to prioritize when instructions conflict. \\

\midrule
\categoryrow
\multicolumn{2}{c}{\textbf{Usecase (32.8\%)}} \\
Common Patterns (20.2\%) & Common patterns that can be reused by the agent. \\
Examples (17.7\%) & Example usecases of the skill. \\

\midrule
\categoryrow
\multicolumn{2}{c}{\textbf{Context (28.2\%)}} \\
Ambiguity Handling (10.5\%) & %
Guidance for resolving ambiguous situations.\\
Decision Tree (8.8\%) & %
Decision-making guidance for alternative courses of action.\\
Target System Orientation (5.0\%) & %
Description of the target system or environment.\\
Guardrail Condition (4.2\%) & %
Constraints that prevent undesirable agent behavior.\\
Additional Context (2.9\%) & %
Guidance on when to retrieve additional context.\\
Persona Injection (1.3\%) & %
Instructions for assigning a persona to the agent.\\

\midrule
\categoryrow
\multicolumn{2}{c}{\textbf{Output Format (23.5\%)}} \\
Return Artifact (21.9\%) & %
Instructions for producing the final output artifact. \\
Lessons Learned (1.3\%) & Recording useful lessons for future invocations of the skill. \\
Procedural Documentation (1.3\%) & %
Instructions for documenting the execution procedure.\\

\midrule
\categoryrow
\multicolumn{2}{c}{\textbf{Practice (23.1\%)}} \\
Best Practices (11.8\%) & %
Recommended practices for using the skill.\\
Anti-Patterns (11.3\%) & %
Practices the agent should avoid.\\

\midrule
\categoryrow
\multicolumn{2}{c}{\textbf{Evaluation (21.4\%)}} \\
Review Checklist (12.6\%) & %
Checklist for validating results before completion.\\
Running Tests (9.2\%) & Instruction on how to run tests. \\
Debugging (1.3\%) & %
Guidance for debugging failed test cases.\\

\midrule
\categoryrow
\multicolumn{2}{c}{\textbf{Tools (4.6\%)}} \\
Tools (4.6\%) & List of preauthorized tools with description. \\

\midrule
\categoryrow
\multicolumn{2}{c}{\textbf{MCP Integration (3.8\%)}} \\
MCP Integration (3.8\%) & Guideline for interacting with MCP Servers. \\

\midrule
\categoryrow
\multicolumn{2}{c}{\textbf{Agent Architecture (2.1\%)}} \\
Agent Composition (1.7\%) & %
Guidance for composing multi-agent systems.\\
Model Choice (0.8\%) & %
Guidance for selecting an appropriate LLM.\\

\midrule
\categoryrow
\multicolumn{2}{c}{\textbf{Other (5.9\%)}} \\
Skill in a Skill (1.7\%) & References to nested skill. \\
Metadata (1.7\%) & Metadata describing the skill. \\
Explaining to Human (1.3\%) & Instructions for communicating results to human users. \\

\bottomrule
\end{tabular}
}
\end{table*}

We identified a hierarchical taxonomy comprising 13 higher-level and 44 lower-level semantic components from 224 \texttt{SKILL.md} files containing 1,615 H2-level sections. Table~\ref{table:body-semantic-component} presents the identified components, their definitions, and their frequency of occurrence. The three most common higher-level components are \textit{Task} (74.4\%), \textit{Introduction} (63.5\%), and \textit{References} (57.1\%). \textit{Task} primarily captures procedural guidance, such as step-by-step instructions, subtasks, and error handling, enabling agents to accomplish the intended objective. \textit{Introduction} provides contextual information, including skill summaries, prerequisites, and triggering conditions that help agents determine when and how to apply the skill. \textit{References} complements these components by pointing agents to supporting resources, such as reference documents, utility scripts, commands, and domain knowledge. Together, these components form the core structure of most real-world \texttt{SKILL.md} files.

\begin{boxH}
\textbf{Finding 1:} Although \texttt{SKILL.md} files are free-form by design, they follow some structure. \textit{Task}, \textit{Introduction}, and \textit{References} form the backbone of most skills, defining what the agent should do, when to do it, and where to find additional guidance.

\end{boxH}

\subsection*{\textbf{RQ2: \skillmd Best Practices \& Skill Smells}} \label{results:rq2}

Our MLR of 29 online resources identified 26 best practices for authoring \texttt{SKILL.md} files. Mapping these best practices to the semantic components identified in RQ1 (Section~\ref{results:rq1}) revealed that only 7 of the 13 higher-level semantic components are governed by explicit authoring guidance. The remaining six components \textit{Introduction}, \textit{Evaluation}, \textit{MCP Integration}, \textit{Agent Architecture}, \textit{Tools}, and \textit{Other} are not covered by any identified best practice. Additionally, we found that 8 of the 26 best practices apply to \texttt{SKILL.md} at a level above individual semantic components: 6 target the frontmatter, which follows a structured schema \cite{skill-specification}, and 2 apply to the \texttt{SKILL.md} body as a whole. 
For each component, Table~\ref{table:best-practices-semantic-components} reports the number of associated best practices and the total number of sources highlighting those best practices (summed across the component's best practices). Together, these results show that existing guidance covers only part of the semantic structure of \texttt{SKILL.md} files, leaving several commonly occurring components without explicit authoring recommendations. The complete list of best practices and their supporting sources is available in the supplementary material~\cite{SupplementaryMaterials}.

\begin{boxH}
\textbf{Finding 2:} Existing \texttt{SKILL.md} authoring guidance is incomplete. Only 7 of the 13 semantic components identified in practice are governed by explicit best practices, leaving nearly half of the semantic structure without documented authoring recommendations.
\end{boxH}

\begin{table}[t!]
\caption{Distribution of Best Practices for Semantic Components}
\label{table:best-practices-semantic-components}
\centering
\resizebox{0.95\columnwidth}{!}{
\begin{tabular}{llcc}
\toprule
\textbf{Scope} & \textbf{Semantic Component} & \textbf{\# Best Practices} & \textbf{\# Sources} \\
\midrule
Frontmatter & - & 6 & 20 \\
\midrule
\multirow{13}{*}{\skillmd Body} & References & 3 & 15 \\
& Context & 5 & 12 \\
& Body as a Whole & 2 & 9 \\
& Principles & 6 & 10 \\
& Task & 1 & 16 \\
& Output Format & 1 & 11 \\
& Usecase & 1 & 6 \\
& Practice & 1 & 4 \\
& Introduction & 0 & - \\
& Evaluation & 0 & - \\
& Tools & 0 & - \\
& MCP Integration & 0 & - \\
& Agent Architecture & 0 & - \\
& Other & 0 & - \\
\bottomrule
\end{tabular}
}
\end{table}

Building on these 26 best practices, we derived a taxonomy of corresponding violations for \skillmd files, which we term \emph{skill smells}. 
Table~\ref{table:skill-smells} presents the resulting \textit{Skill Smells}, their acronyms, definitions, and grouping into categories based on the nature of the underlying deficiency.

\begin{figure}[t!]
  \centering
  \includegraphics[width=0.85\columnwidth]{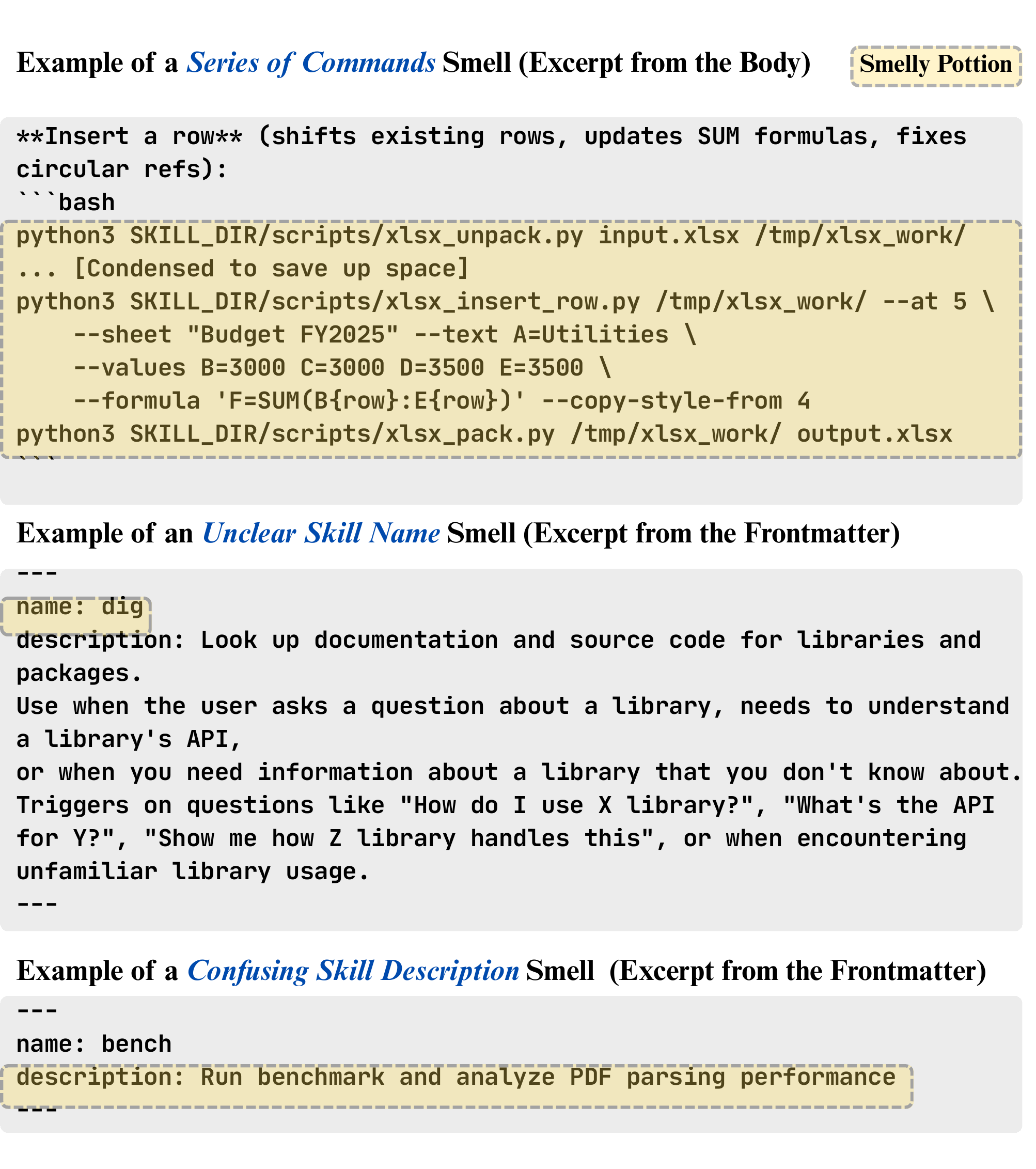}
  \caption{Example of Skill Smells
  }
  \label{fig:rq3-smell-examples}
\end{figure}

Figure~\ref{fig:rq3-smell-examples} presents three representative examples of skill smells identified in \skillmd files. We present the three most prevalent presence-based smells among the six identified presence-based skill smells. Absence-based smells (e.g., \textit{No Validation Step}) are excluded from this illustrative analysis, as their defining characteristic (omission of expected information) renders them unsuitable for representation through discrete code excerpts. In our results we have 20 absence-based smells. The \textit{Series of Commands} smell is shown in a body excerpt that prescribes a rigid, line-by-line sequence of shell commands, including hardcoded file paths and arguments, rather than describing the underlying objective and allowing the agent to determine the appropriate invocation. The \textit{Unclear Skill Name} smell is shown in a frontmatter excerpt where the skill is named \texttt{dig}, a name that conveys no information about the skill's purpose and could plausibly refer to many unrelated tasks. Finally, the \textit{Confusing Skill Description} smell is shown in a description that states only that the skill runs a benchmark and analyzes performance, without indicating \textit{when} the skill should be invoked or \textit{what} kind of input it expects, leaving the agent unable to reliably determine its relevance to a given task. Due to space limitations, examples for the remaining skill smells can be found in the supplementary material \cite{SupplementaryMaterials}.

\definecolor{presence}{rgb}{0.90,0.90,0.90}

\begin{table*}[t!]
\caption{Taxonomy of skill smells, grouped by the underlying type of quality deficiency. Skills smells that are presence-based are \hlc[presence]{highlighted}.}
\label{table:skill-smells}
\centering
\resizebox{\textwidth}{!}{%
\begin{tabular}{m{4cm}m{2cm}p{12cm}}
\toprule
\textbf{Skill Smell Name} & \textbf{Acronym} & \textbf{Definition} \\
\midrule
\categoryrow
\multicolumn{3}{c}{\textbf{\textit{Under-Specified Guidance}}} \\
The Stepless Workflow & TSW & \skillmd %
describes an entire workflow as a single block of prose instead of decomposing it into steps.
\\
\rowcolor{presence} The Option Buffet & TOB & \skillmd %
provides multiple alternative tools or libraries without recommending a default choice.\\
Missing Utility Script & MUS & %

\skillmd omits utility scripts for tasks that are better handled with scripts.
\\

\multirow{2}{*}{Missing Decision Tree} & \multirow{2}{*}{MDT} & \skillmd does not provide a decision tree to assist the agent in choosing the right approach based on the situation. \\
\categoryrow
\multicolumn{3}{c}{\textbf{\textit{Over-Prescribed Guidance}}} \\
\rowcolor{presence} \multirow{2}{*}{Series of Commands} & \multirow{2}{*}{SOC} & Prescribes exactly which steps to run and in which order, instead of allowing the agent to adapt its execution.\\
\categoryrow

\multicolumn{3}{c}{\textbf{\textit{Missing Verification \& Feedback Loop}}} \\
No Validation Step & NVS & \skillmd treats output generation as a one-shot process without any validation loops. \\

\multirow{2}{*}{Execute Without a Plan} & \multirow{2}{*}{EWP} & %

\skillmd directs the agent to execute complex tasks without an intermediate planning or validation stage. \\

Never Asks Human & NAH & \skillmd do not provide a mechanism for the agent to request human feedback. \\

\categoryrow
\multicolumn{3}{c}{\textbf{\textit{Missing Follow-Through Guards}}} \\
Rationalization Loophole & RL & %
\skillmd provides no guidance to discourage the agent from rationalizing or skipping required steps. \\
No Progress Tracking & NPT & \skillmd requires a multi-step workflow, but does not provide a mechanism to track progress. \\
\categoryrow
\multicolumn{3}{c}{\textbf{\textit{Context Bloat}}} \\
\multirow{2}{*}{Undelegated Detail} & \multirow{2}{*}{UD} & %
\skillmd embeds low-level implementation details instead of delegating them to reference documents or scripts. \\
Lengthy Skill Body & LSB & \skillmd body exceeds the recommended 5,000 words by best practices \cite{claude-api-docs, microsoft-learn, perplexity-skills, bibek-poudel, gemini-cli-skills, claude-skills-hub}.\\
Lengthy Skill Name & LSN & Frontmatter's name field exceeds the recommended 64 characters by best practices \cite{claude-api-docs, microsoft-learn, perplexity-skills, bibek-poudel, gemini-cli-skills, claude-skills-hub}. \\
Lengthy Skill Description & LSD & Frontmatter's description field exceeds the recommended 1,024 characters by best practices \cite{claude-api-docs, microsoft-learn, perplexity-skills, bibek-poudel, gemini-cli-skills, claude-skills-hub}. \\
\rowcolor{presence} \multirow{2}{*}{Confusing Skill Description} & \multirow{2}{*}{CSD} & Frontmatter's description field should have a structure of [What it does] + [When to use it] + [Keywords]. However, the \skillmd frontmatter's description field fails to provide at least one of them. \\

\categoryrow
\multicolumn{3}{c}{\textbf{\textit{Missing Safeguards}}} \\

\multirow{2}{*}{No Guardrails} & \multirow{2}{*}{NG} & \skillmd does not provide any guardrails to prevent the agent from attempting an inappropriate or impossible task. \\

\multirow{2}{*}{Buried Gotchas} & \multirow{2}{*}{BG} & %

\skillmd fails to highlight critical warnings or caveats that the agent should not overlook using recommended gotcha headers. \\

Missing Usage Rules & MUR & %
\skillmd omits rules governing when or how the skill should be used.
\\

Missing Caveats & MC & %
\skillmd omits common caveats and their resolution.
\\

\categoryrow
\multicolumn{3}{c}{\textbf{\textit{Inadequate Contextual Grounding}}} \\
Missing Example & ME & \skillmd do not provide examples that can assist the agent in obtaining sufficient context. \\
\multirow{2}{*}{Time Sensitive Skill} & \multirow{2}{*}{TSS} & \skillmd contains time-sensitive information that requires the agent to know the current time and it becomes outdated after a certain point. \ \\

\categoryrow

\multicolumn{3}{c}{\textbf{\textit{Security Hazard}}} \\
XML Included Description & XID & Frontmatter's description contains XML tags which can inject unintended instructions into prompt \\

\categoryrow
\multicolumn{3}{c}{\textbf{\textit{Convention \& Style Violations}}} \\
\rowcolor{presence} \multirow{2}{*}{Backslash Path} & \multirow{2}{*}{BP} & \skillmd contain paths denoted using a backslash. Agents navigate the skill directory like a file system so paths must be written using a frontslash. \\
\rowcolor{presence} Unclear Skill Name & USN & %
Uses a skill name that does not clearly convey the skill's capability or action.
\\
\rowcolor{presence} \multirow{2}{*}{Non Third Person Description} & \multirow{2}{*}{NTPD} & Frontmatter's description field should always be written in third person, but the description field is not. As inconsistent point-of-view can cause discovery problems. \\

\categoryrow
\multicolumn{3}{c}{\textbf{\textit{Unstructured Output}}} \\
\multirow{2}{*}{Missing Template} & \multirow{2}{*}{MT} & \skillmd does not provide a template even though the agent needs to produce an output in a specific format. \\
\bottomrule
\end{tabular}
}
\end{table*}

\subsection*{\textbf{RQ3: Prevalence of Skill Smells}} \label{results:rq3}
Due to the large number of skill smells in our catalog, manually inspecting each \skillmd file for the presence of every smell was infeasible. We therefore developed an automated approach to detect skill smells in a given \skillmd file. We adopted a hybrid detection strategy. Smells that can be identified without semantic reasoning, such as those defined by the count of words or the presence of specific characters, are detected through static analysis. The smells that require contextual understanding, on the other hand, are detected using an LLM-based approach. The LLM-based detector frames each smell as a binary classification task, returning \emph{Yes} when the smell is present and \emph{No} otherwise. As justified in Section~\ref{method:rq3}, we ran our tool with a 4-bit AWQ quantized variant of Qwen3.6-27B and conducted an evaluation on the manually labeled dataset (Table~\ref{table:skill-smell-freq}). Our evaluation reports a weighted F1 score of $0.78$, a weighted precision score of $0.79$, and a weighted recall score of $0.78$, providing sufficient accuracy for large-scale prevalence analysis.

\begin{figure}[b!]
  \centering
  \includegraphics[width=1.0\columnwidth]{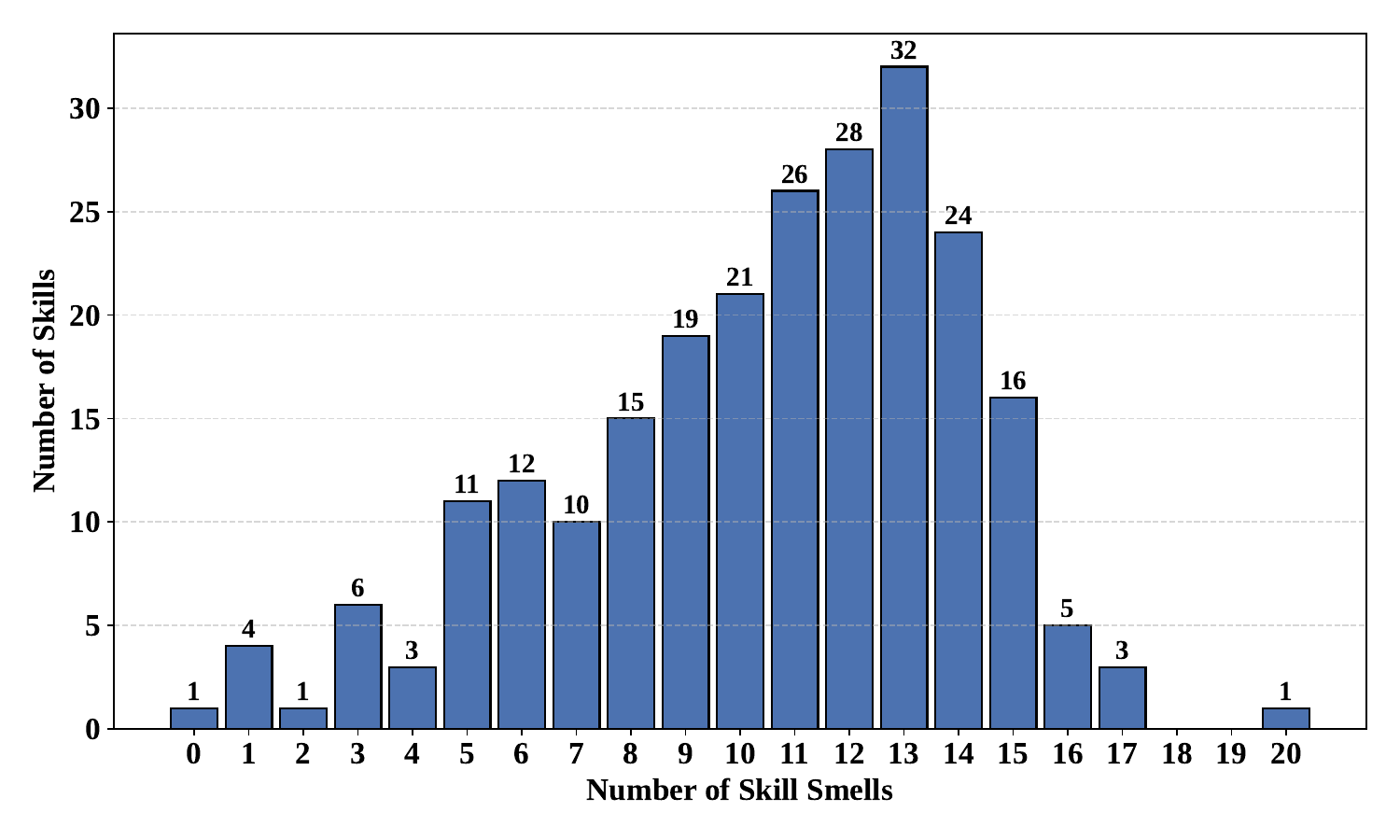}
  \caption{Distribution of the Number of Skill Smells per \skillmd File}
  \label{fig:rq3-smell-histogram}
\end{figure}

We next applied the detector to all 238 \texttt{SKILL.md} files to quantify skill smell prevalence. Specifically, we measured both the prevalence of each individual skill smell across the corpus and the number of co-occurring skill smells within each \texttt{SKILL.md} file.

\begin{table}[t!]
\caption{Distribution of Skill Smells in the Groundtruth Dataset Used to Evaluate SSD and the Entire Skills Sample.  \hlc[presence]{Highlighted} rows demonstrate \textit{Statically Detectable} smells. Therefore, they were absent in the Validation Dataset used to evaluate the accuracy of detection of \textit{Semantically Detectable} smells.}
\label{table:skill-smell-freq}
\centering
\resizebox{0.9\columnwidth}{!}{%
\begin{tabular}{ccc}
\toprule
\textbf{Skill Smell} & \multicolumn{2}{c}{\textbf{Frequency of Smell}}  \\
& Validation Dataset (53 Skills) & Entire Sample (228 Skills) \\
\midrule
RL & 52 (98\%) & 223 (94\%) \\
BG & 30 (57\%) & 192 (81\%) \\
EWP & 35 (66.0\%) & 185 (78\%) \\
NAH & 38 (72\%) & 184 (77\%) \\
NPT & 47 (89\%) & 169 (71\%) \\
MC & 30 (57\%) & 170 (71\%) \\
NVS & 43 (81\%) & 165 (69\%) \\
MDT & 34 (64\%) & 165 (69\%) \\
NG & 28 (53\%) & 159 (67\%) \\
MUS & 37 (70\%) & 156 (66\%) \\
SOC & 32 (60\%) & 148 (62\%) \\
UD & 18 (34\%) & 109 (46\%) \\
MUR & 12 (23\%) & 105 (44\%) \\
ME & 17 (32\%) & 80 (34\%) \\
TSW & 24 (45\%) & 75 (32\%) \\
CSD & 18 (34\%) & 75 (32\%) \\
MT & 30 (57\%) & 60 (25\%) \\
USN & 12 (23\%) & 41 (17\%) \\
TOB & 19 (36\%) & 19 (8\%) \\
TSS & 2 (4\%) & 6 (3\%) \\
NTPD & 2 (4\%) & 8 (3\%) \\
\rowcolor{presence} BP & - & 3 (1\%) \\
\rowcolor{presence} LSB & - & 1 ($\approx 0\%$) \\
\rowcolor{presence} LSD & - & 1 ($\approx 0\%$) \\
\rowcolor{presence} LSN & - & 0 (0\%) \\
\rowcolor{presence} XID & - & 0 (0\%) \\

\bottomrule
\end{tabular}
}
\end{table}

Table~\ref{table:skill-smell-freq} summarizes the prevalence of individual skill smells. Overall, skill smells are highly pervasive: 11 of the 26 identified skill smells occur in more than 50\% of the analyzed \texttt{SKILL.md} files. The most prevalent smell is \textit{Rationalization Loophole (RL)}, which appears in 94\% of skills, whereas \textit{Lengthy Skill Name (LSN)} and \textit{Lengthy Skill Description (LSD)} are the least common. Figure~\ref{fig:rq3-smell-histogram} further shows that skill smells rarely occur in isolation. Only one of the 238 analyzed \texttt{SKILL.md} files are entirely free of skill smells, and skills contain an average of $10.5$ skill smells each.

\begin{boxH}
\textbf{Finding 3:} Skill smells are pervasive in real-world \texttt{SKILL.md} files. 11 of the 26 identified skill smells appear in more than half of the analyzed skills, only one skill is entirely smell-free, and each \texttt{SKILL.md} contains an average of 10.5 skill smells. This indicates violations of authoring best practices are therefore the norm rather than the exception.
\end{boxH}

\subsection*{\textbf{RQ4: How Skill Smell Prevalence Evolves Over Time}} \label{results:rq4}

To examine whether skill smells are corrected or persist over time, we conducted a longitudinal analysis of their weekly prevalence using the methodology described in Section \ref{method:rq4}. For each week, we computed the proportion of existing skills exhibiting a given smell, where a skill's state is taken from its most recent commit (last-observation carried forward).

\begin{figure}[t!]
  \centering
  \includegraphics[width=1.0\columnwidth]{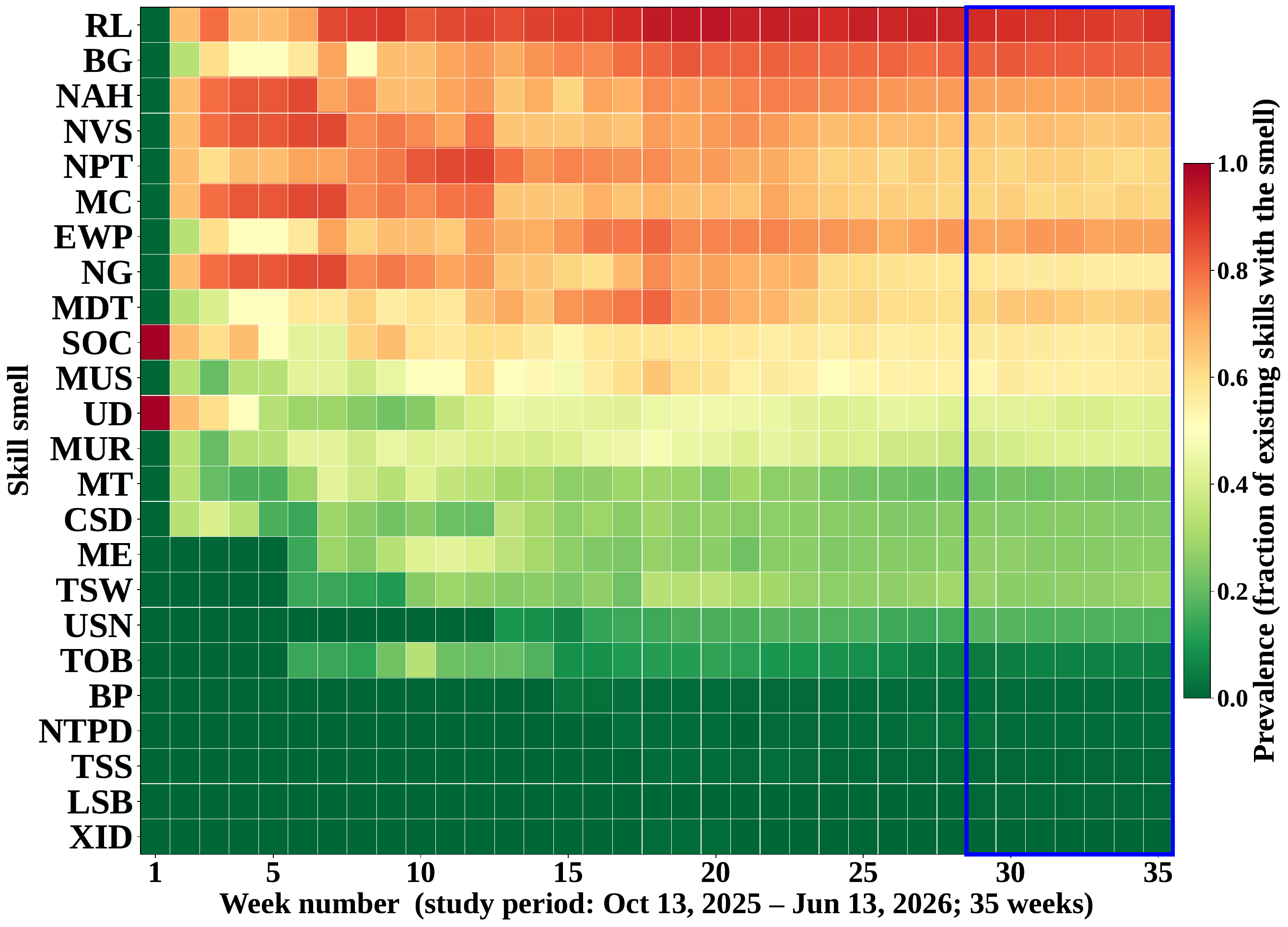}
  \caption{Temporal prevalence of skill smells across all existing skills, by week (green = 0\% of existing skills affected, red = 100\%). The area surrounded with a blue rectangle demonstrates the weeks where all \skillmd files with at least 2 commits (142) were present.
  }
  \label{fig:rq4-smell-heatmap}
\end{figure}

We present the results into a smell-by-week matrix and rendered it as the heatmap in Figure~\ref{fig:rq4-smell-heatmap}, in which each row corresponds to a skill smell, each column to a week, and each cell is colored on a continuous scale from green (a prevalence of 0; no existing file exhibits the smell) to red (a prevalence of 1; every existing file exhibits it). Because \skillmd files enter the corpus over time, the number of files underlying each weekly column is not constant, growing from a single file in the first week to the full set of 142 (Week 29 onwards). The specific number of \skillmd included, and the prevalence of each skill smell per week, is available in the supplementary material~\cite{SupplementaryMaterials}.

Two smells that had zero prevalence across all weeks are omitted from the figure. \textit{Lengthy Skill Name} never appears in our dataset. \textit{Lengthy Skill Description} does appear (Table~\ref{table:skill-smell-freq}), but only in a single \skillmd with one commit; per our RQ4 methodology (Section~\ref{method:rq4}), we exclude all single-commit \skillmd files.

Across all analyzed skill smells, we observe no evidence that their prevalence decreases over time. Instead, the prevalence of each skill smell tends to worsen as time passes.

We acknowledge that the full 35-week window does not apply uniformly to every skill smell, as 51\% of the skills did not yet exist within the first $18$ weeks. 
To account for this, we outline the final block of weeks with a blue rectangle on our figure, by which every \skillmd has been included. Within this region, we observed that the prevalence of each smell remains stable; skill smells are neither addressed nor newly introduced, indicating that skill smells are seldom corrected once introduced and instead tend to persist throughout the development cycle.

To validate that this trend is not driven solely by aggregate statistics, we also examined the evolution of each repository individually. Due to space constraints, we could not include all 142 individual skill-level figures in the paper; instead, they are provided in our supplementary material~\cite{SupplementaryMaterials}. Across these figures, we observed a similar pattern of once introduced, skill smells are seldom removed during subsequent development.%

\begin{boxH}

\textbf{Finding 4:} Skill smells rarely disappear once introduced. Across both repository-level and individual-skill analyses, we find no systematic evidence that skill smells are corrected over time, suggesting they persist throughout the evolution of \texttt{SKILL.md} files.

\end{boxH}

%% file: Sections/implications.tex
Our results show that skill smells are pervasive in real-world \texttt{SKILL.md} files and persist throughout repository evolution rather than being corrected over time. The direct impact of individual skill smells on agent performance remains an open question. Since \texttt{SKILL.md} files become part of an agent's context window~\cite{liu2026agentskillswildempirical}, we therefore draw on empirical findings from the broader LLM and agent literature, arguing that evidence on analogous context-level deficiencies provides a reasonable proxy for understanding how skill smells may influence agent behavior, efficiency, robustness, and safety.

Authoring high-quality \texttt{SKILL.md} files are fundamentally a balancing act between three complementary design principles: providing the right amount of procedural guidance, enabling effective verification, and supplying high-quality context. First, \textit{Under-Specified Guidance} and \textit{Over-Prescribed Guidance} represent opposite ends of the same spectrum. Skills should provide sufficient procedural structure to help agents decompose tasks and make informed decisions, yet avoid prescribing execution so rigidly that agents lose the flexibility to adapt when circumstances change. Prior work similarly shows that explicit workflow decomposition and decision-support mechanisms improve agent performance, whereas excessive prompt specification reduces an LLM's ability to follow instructions reliably~\cite{hsiao2025proceduralknowledgeimprovesagentic,huang2024affectsstabilitytoollearning,zhai2025aaai,yang2025prompts}. 

Second, high-quality skills should support the execution process through planning, validation, feedback, progress tracking, and operational safeguards. These mechanisms enable agents to verify intermediate results, recognize infeasible tasks, and prevent errors from propagating to the final output. Prior work consistently demonstrates that planning, self-refinement, structured checklists, and runtime guardrails improve reliability, reduce unsafe behavior, and enhance task performance~\cite{wang2023plan,madaan2023self,chae2026web,thaman2026reward,zhang2024recognizing,wang2025agentspec}. \textit{Missing Verification \& Feedback Loops}, \textit{Missing Follow-Through Guards}, and \textit{Missing Safeguards} represent skill smells in this category.

Finally, effective skills should provide context that is both sufficient and focused. \textit{Context Bloat} and \textit{Inadequate Contextual Grounding} again represent opposing failure modes: unnecessary information increases reasoning burden and competes for limited context-window capacity, while insufficient grounding forces agents to compensate through inference. Prior work shows that both irrelevant context and missing contextual information degrade reasoning quality, highlighting that effective \texttt{SKILL.md} files should provide the right information at the appropriate level of detail rather than simply more or less information~\cite{levy2024same,hsieh2024ruler,wang-etal-2022-super,tan2023towards}. Collectively, these findings suggest that many skill smells arise not because a skill contains too much or too little of a particular element, but because it fails to strike the balance required for reliable and efficient agent execution.

Skill smells can also have security implications. \textit{Security Hazard} skill smells expose agents to prompt-injection attacks by embedding content that becomes part of the agent's context. Prior work shows that LLMs are susceptible to prompt injection through untrusted retrieved content~\cite{greshake2023not}, making structured tags in \texttt{SKILL.md} files a potential attack vector. These smells, therefore, increase the attack surface of Agent Skills and may compromise the security of agentic systems.

Seemingly superficial skill smells, such as those in the \textit{Convention \& Style Violation} and \textit{Unstructured Output} categories, primarily affect naming conventions, metadata, or output templates, yet can substantially affect agent behavior. Tool names, descriptions, and parameter schemas are among the primary signals LLMs use when selecting tools~\cite{sneh2025tooltweak}, while explicit output templates improve adherence to user-defined output requirements~\cite{mao2025prompts}. Consequently, inconsistent metadata or missing output templates can hinder skill discovery, reduce correct skill invocation, and produce inconsistent outputs. These findings suggest that authoring conventions are more than stylistic guidelines; they directly shape how agents discover, invoke, and interact with skills.

Our findings have implications for researchers, tool builders, and practitioners. We introduce skill smells as a new software quality concern for agentic systems, opening several research directions. Future work should quantify the causal impact of individual skill smells on agent performance, investigate interactions among multiple smells, evaluate their effects across different foundation models and agent architectures, and study how skill smells influence increasingly complex multi-agent workflows.

Our findings also suggest that the high prevalence and persistence of skill smells make manual quality control unlikely to be sufficient. Our taxonomy provides a foundation for automated quality-assurance tools for Agent Skills. Similar to linters and static analyzers for source code, future tools could utilize the results of SSD to recommend refactorings, automatically repair common issues, and be integrated into IDEs, CI pipelines, and Agent Skill marketplaces. %

Finally, developers should treat \texttt{SKILL.md} files as executable software artifacts rather than passive documentation. %
As Agent Skills become increasingly shared across organizations and marketplaces, establishing and adhering to high-quality authoring practices will be essential to building trustworthy and maintainable agent ecosystems.

Taken together, our findings suggest that skill smells should not be viewed as isolated documentation issues, but as an emerging software quality concern for agentic systems. %
This combination of high prevalence, long-term persistence, and plausible downstream consequences suggests that the community has an opportunity to intervene while Agent Skills are still in their early stages of adoption. Software engineering has repeatedly witnessed similar trajectories with software artifacts such as source code, tests, and security configurations, in which quality concerns were initially underestimated, prompting decades of research on code smells, test smells, security smells, automated detection, and refactoring techniques. Our findings suggest that Agent Skills present a similar opportunity for early intervention before poor authoring practices become deeply embedded in the ecosystem.

%% file: Sections/threats.tex
\textbf{Construct Validity.} One threat concerns the subjectivity of manually identifying semantic components and labeling skill smells. We mitigated this through using independent coding by two authors, followed by negotiated agreement to resolve all disagreements~\cite{corbin2015basics,strauss1998basics}. Another unavoidable threat is that our best practices are derived from grey literature rather than peer-reviewed guidelines. To reduce potential bias, we independently reviewed the top 50 retrieved sources and retained only practices supported by a strict majority of the selected sources. A further threat concerns the reliability of our LLM-based detector for semantically detectable skill smells, which we mitigated by validating it on a manually curated ground-truth dataset. %

\textbf{External Validity.} A key threat to external validity is that our sample was drawn exclusively from the \texttt{skills.sh} marketplace~\cite{skills-sh}. Furthermore, restricting the sample to popular English-language skills may limit generalizability and could underestimate the prevalence of skill smells in the broader ecosystem. Another threat concerns our use of a single LLM, Qwen3.6-27B (4-bit AWQ), for detecting semantically detectable skill smells; prevalence estimates may differ with other models or quantization schemes. %

%% file: Sections/conclusion.tex
Our study presents the first systematic characterization of \texttt{SKILL.md} files as a software artifact. We identify a taxonomy of 13 higher-level and 44 lower-level semantic components, derive 26 best practices, and introduce \textit{skill smells} as violations of these practices. Our empirical study shows that skill smells are highly prevalent and become more common as skills evolve over time. These findings establish \texttt{SKILL.md} quality as an important SE concern and motivate automated support for refactoring skill smells during skill authoring. All code, data, and replication materials are available in the supplementary material~\cite{SupplementaryMaterials}.